\documentclass{article}%
\usepackage{amsfonts}
\usepackage{amsmath}
\usepackage{amssymb}
\usepackage{graphicx}%
\providecommand{\U}[1]{\protect\rule{.1in}{.1in}}
\newtheorem{theorem}{Theorem}
\newtheorem{acknowledgement}[theorem]{Acknowledgement}

\begin{document}

\title{The quartic oscillator in an external field and the statistical physics of
highly anisotropic solids}
\author{Victor Barsan\\IFIN-HH, 407 Atomistilor Street, Magurele-Bucharest 077125, \\Romania; vbarsan@theory.nipne.ro}
\maketitle

\begin{abstract}
The statistical mechanics of 1D and 2D Ginzburg-Landau systems is evaluated
analytically, via the transfer matrix method, using an expression of the
ground state energy of the quartic anharmonic oscillator in an external field.
In the 2D case, the critical temperature of the order/disorder phase
transition is expressed as a Lambert function of the inverse inter-chain
coupling constant.

\end{abstract}

\section{Introduction}

The physics of 1D Ginzburg-Landau systems, described by a polynomial
Hamiltonian, with quartic anharmonicities, is of utmost interest, for several
domains, from quantum field theory to structural phase transitions. 2- or 3D
aggregates of such chains describe anisotropic anharmonic solids. Examples of
this kind of 2D systems are the ultra-thin ferroelectric films, obtained quite
recently \cite{Ducharme1998}. In ultra-thin polymeric ferroelectric films, the
strong dipols constituting the polymer chain allow the manipulation of the
individual monomer by a local field, making this class of materials very
attractive for molecular electronics \cite{Cai2004}[Cai2004].

Theoretically, one of the most popular approaches to the study of
Ginzburg-Landau anisotropic systems is the transfer matrix method. With this
method, the evaluation of the free energy of a $\left(  N+1\right)  $ -
dimensional classical system is equivalent to the evaluation of the ground
state energy of a quantum $N$ - dimensional system, described by an neffective
Schrodinger equation. The classical $\left(  N+1\right)  $ - dimensional
system and its quantum $N$ - dimensional counterpart are sometimes termed dual
systems \cite{vb/pm/2008}. Any progress in understanding the physics of such a
system is mirrorede by a progress in understanding its dual system. In
traditional approaches \cite{SSF}, \cite{KS}, the solution of the quantum
problem is used in order to find its classical counterpart. However, in some
Monte Carlo simulations \cite{wang}, the classical 2D system was studied in
order to solve the quantum 1D system.

Technically speaking, the transfer matrix approach, as developed by Scalapino
and his co-workers \cite{SSF}, \cite{SS} is mainly a convenient method of
evaluating the partition sum, expressed as a functional integral. The first
decade of investigations of this subject - until the early '80s - was
dominated by significant analytical results. For the next two decades, the
efforts were mainly focused on numerical investigations. However, in the
recent years, some exact results, or at least analytical approximations,
obtained in the study of the quartic oscillator, show a change of this
tendency. They allow us to give analytical responses to some important
problems, which have been received, previously, only numerical solutions.

To this trend belongs also the present paper. We use an analytical
approximation of the ground state energy of a quartic oscillator in an
external field, obtained by Van der Straeten and Naudts \cite{Van}, in order
to find the exact statistical mechanics of a chain of classical anharmonic
oscillators, and the mean-field approximation of the statistical mechanics of
a planar array of coupled chains.

In fact, the formula proposed in \cite{Van} for the ground state energy of a
quartic oscillator in an electric field is merely an ansatz than a rigorously
obtained result. Also, it depends only on the external field - any dependence
of temperature of the effective mass of the oscillator, essential for the
description of the statistical mechanics of the linear chain, is lost, being
replaced by some numerical parameters. In our approach, both these drawbacks
of the Van der Straeten-Naudts solution are removed. The temperature
dependence is restored using a comparison with similar results obtained for
the 1D Ising model in external field. The adequacy of the ansatz is checked
and confirmed by the physical character of its predictions and, more than
this, by the fact that it gives results obtained using alternative methods.
This is an interesting example of the fact that, while studying dual systems,
a progress made in understanding one of them is mirrored in the understanding
of the other.

The outline of this paper is as follows. In section 2, we shall sketch the
general frame of the problem, at least for sake of clearly defining the
notations. In Section 3, we shall expose in some details the Van der
Straeten-Naudts solution, in order to properly understand its advantages and
drowbacks. In Section 4, this solution, with restored temperature depedence,
is used in order to find the electrical polarization of the anharmonic chain.
Section 5 is devoted to the physics of 2- and 3D arrays of such chains,
treated in the mean-field approximation. A new analytic formula, for the
critical temperature of the order/disorder transition, as a function of the
inter-chain coupling, is obtained. Its small coupling limit coincides with a
formula obtained in \cite{scal.im.pinc}, confirmin the adequacy of the ansatz
contained in the Van der Straeten-Naudts solution. The last section is devoted
to conclusions.

\bigskip

\section{Chains of anharmonic oscillators}

Basically, the physical system to be studied in this paper is a chain of
classical anharmonic oscillators, described by the Hamiltonian:%

\begin{equation}
H_{cl}=\sum_{i=1}^{N}\frac{1}{2}m_{i}\left(  \frac{du_{i}}{dt}\right)
^{2}+\sum_{i=1}^{N}\left[  \left(  \frac{1}{2}Au_{i}^{2}+\frac{1}{4}Bu_{i}%
^{4}\right)  +\sum_{j=1}^{N}\frac{1}{2}c_{ij}\left(  u_{i}-u_{j}\right)
^{2}\right]  \label{1}%
\end{equation}
Each oscillator is an atom; $i,j$ indicate lattice sites (equilibrum positions
of atoms); $u_{i}$ - displacements of displacing atoms with respect to some
heavy ions or reference lattice. Periodic boundary conditions are assumed. The
coefficient $A$ is defined by the attractive interactions of the mobile atom
with the reference lattice, $B$ - by short-range repulsive interactions, and
$c_{ij}$ - by elastic interactions between displacing atoms. If (1) describes
a lattice which is unstable against a displacive transition,
$A<0,\ B>0,\ c_{ij}>0.$ In this situation, the potential energy on site
appearing in (1) has two minima, at%

\begin{equation}
u_{0}=\left(  \frac{\left\vert A\right\vert }{B}\right)  ^{1/2} \label{2}%
\end{equation}

We shall replace the discrete Hamiltonian (1) with a continuum representation:%

\begin{equation}
H_{cl}=\int\frac{dx}{l}\left[  \frac{1}{2m}p\left(  x\right)  ^{2}+\frac{A}%
{2}u\left(  x\right)  ^{2}+\frac{B}{4}u\left(  x\right)  ^{4}+\frac{1}%
{2}mc_{0}^{2}\left(  \frac{du}{dx}\right)  ^{2}\right]  \label{3}%
\end{equation}
where $l$ is the lattice spacing, $c_{0}$ - the sound velocity, and $x_{j}=jl
$ locates an atom in the continuum representation.

The equilibrum thermodynamics of this 1D model can be obtained from the
classical partition function:%

\begin{equation}
Z=\int\mathcal{D}u\mathcal{D}pe^{-\beta\mathcal{H}\left(  u,p\right)  }
\label{4}%
\end{equation}
where, as usual,%

\begin{equation}
Z_{p}=\left(  2\pi k_{B}T\right)  ^{N/2} \label{5}%
\end{equation}
and%

\begin{equation}
Z_{u}=e^{-N\beta\epsilon_{0}} \label{6}%
\end{equation}
with $\epsilon_{0}$ - the ground state energy of the Schrodinger equation:%

\begin{equation}
\left(  -\frac{1}{2m^{\ast}}\frac{d^{2}}{du^{2}}+\frac{A}{2}u^{2}+\frac{B}%
{4}u^{4}\right)  \Psi_{n}\left(  u\right)  =\epsilon_{n}\Psi_{n}\left(
u\right)  \label{7}%
\end{equation}

\begin{equation}
m^{\ast}=m\frac{c_{0}^{2}}{l^{2}k_{B}^{2}T^{2}} \label{8}%
\end{equation}

The free energy $F$ of the chain is:%

\begin{equation}
F=-k_{B}T\ln Z_{p}Z_{u}=F_{p}+F_{u} \label{9}%
\end{equation}
but, in fact, only the "interaction term", $F_{u},$ produces interesting
physical effects. The free energy per particle is:%

\begin{equation}
f_{u}=\epsilon_{0} \label{10}%
\end{equation}

In the presence of an electric field, a term%

\[
-\mathcal{E}qu_{i}=-pu_{i}
\]
is added to the on-site energy (the notation used in the r.h.s. intends to
compromize with that used in \cite{Van}), and the quantum Hamiltonian (7) becomes:%

\begin{equation}
H_{q}=-\frac{1}{2m^{\ast}}\frac{d^{2}}{du^{2}}+\frac{A}{2}u^{2}+\frac{B}%
{4}u^{4}-pu \label{11}%
\end{equation}

So, the average value of the displacement can be written as:%

\begin{equation}
\left\langle u\right\rangle =-\frac{\partial\epsilon_{0}\left(  p\right)
}{\partial p} \label{12}%
\end{equation}
where $\epsilon_{0}\left(  p\right)  $ is the ground state energy of $H_{q}.$
Also, using any of the Hamiltonians (7), (11), we can compute the average
value of other powers of the order parameter:%

\[
\left\langle u^{2}\right\rangle =-2\frac{\partial\epsilon_{0}}{\partial
\left\vert A\right\vert },\ \ \ \left\langle u^{4}\right\rangle =4\frac
{\partial\epsilon_{0}}{\partial B}
\]

\bigskip The physics of such a system is quite simple; we shall discuss it
starting with the discrete variant, described by the Hamiltonian (1). In the
absence of an external field, each atom oscillates near one of the two minima,
given by (2); the average value of the displacement is zero, at any
temperature. For a planar array of such chains - so, for a 2D system - the
situation is quite different. The inter-chain interaction limits the effect of
fluctuations, and the average displacement along a chain, $\left\langle
u\right\rangle $, can have a non-zero value, if the temperature is lower than
a critical value, $T_{c}$. So, the sistem displays an order/disorder phase
transition, similar to a magnetic/paramagnetic transition, in a magnetic
system. Referring again to the 1D system, an external electric field can
produce a state with $\left\langle u\right\rangle \neq0.$ Also, the effect of
inter-chain interactions, in a 2D system, can be treated, in the mean-field
approximation, as an effective external field, able to trigger a phase
transition. For the description of such a phase transition, it is important to
evaluate the critical temperature as a function of microscopic quantities,
mainly of inter-chain coupling. Also, quantities like the order parameter
$\left\langle u\right\rangle ,$ the polarization and the electric
susceptibility play a similar role to magnetic moment, magnetization and
magnetic susceptibility. In fact, there is an evident connection between the
atomic chain with deep on-site wells and the 1D Ising model, at least for
small external fields.

In principle, the formulae given in this section, mainly (10), (12), can
provide a quite complete description of the "exact" statistical mechanics of
the 1D system (if the eigenvalue $\epsilon_{0}$ is known "exactly"), and a
mean-field variant of such a description, for the 2D system. However, as it is
wellknown, the eigenvalues of (11) cannot be evaluated exactly, even for
$p=0$. Recently, an approximate analytical expression for $\epsilon_{0}\left(
p\right)  $ has been obtained in \cite{Van}. We shall take advantage of this
formula in order to find the polarization, the electrical susceptibility and
other properties of the chain of classical anharmonic oscillators, and of the
2D aggregates of such chains. The outputs of our calculations will be also a
test of the adequacy of the expression of $\epsilon_{0}\left(  p\right)  $
given by \cite{Van}.

In the next section, we shall briefly expose how the expression of
$\epsilon_{0}\left(  p\right)  $ has been obtained, in order to properly
understand its validity and its limitations.

\section{The quantum double well anharmonic oscillator in an external field}

We shall briefly describe the results obtained recently by Van der Straeten
and Naudts \cite{Van} on this subject. The Hamiltonian:%

\begin{equation}
H=\frac{P^{2}}{2m}+\sum_{i=0}^{I}\lambda_{i}Q^{i} \label{13}%
\end{equation}
is written in terms of annihilation and creation operators $a,\ \ a^{+}$ \ of
the ordinary harmonic oscillator with mass $m$ and frequency $\omega_{0}. $ Defining%

\begin{equation}
r^{2}=\frac{\hbar}{m\omega_{0}} \label{14}%
\end{equation}
and using a formula for normal ordering derived in \cite{Jaf}, the Hamiltonian
(1) takes the form:%

\begin{equation}
H=-\frac{\hbar^{2}}{4mr^{2}}\left(  a-a^{+}\right)  ^{2}+\sum_{i=0}^{I}%
\lambda_{i}\left(  \frac{r}{\sqrt{2}}\right)  ^{i}\sum_{k=0}^{\left[  \frac
{i}{2}\right]  }\frac{i!}{2^{k}k!}\sum_{j=0}^{i-2k}\frac{\left(  a^{+}\right)
^{i-2k-j}a^{j}}{j!\left(  i-2k-j\right)  !} \label{15}%
\end{equation}

The expectation value of (15) in a certain state $|\psi_{t}>,$ depending on a
parameter $t$, will be minimized%

\begin{equation}
\frac{\partial}{\partial r^{2}}\left\langle \psi_{t}|H|\psi_{t}\right\rangle
=0 \label{16}%
\end{equation}
in order to obtain an equation which can determine the parameter $t$.

For a quartic oscillator in an external field, $\ \lambda_{1}=-p,\ \lambda
_{2}=\frac{\alpha}{2},\ \lambda_{3}=0,\ \lambda_{4}=\frac{\beta}{4},\ I=4,$
and (16) becomes:%

\begin{equation}
\left(  2t+1\right)  \left(  2\lambda_{2}r_{0}^{4}-\frac{\hbar^{2}}{m}\right)
+6\lambda_{4}r_{0}^{6}\left(  2t^{2}+2t+1\right)  =0 \label{17}%
\end{equation}

If $t$ is known - or, equivalently, if the choice of the state vector
$|\psi_{t}>$ has been done - eq. (17) will give the value of $r_{0};$ this
cubic equation has indeed one real root, if the potential has two wells
$\left(  \alpha<0\right)  .$ The choice of $t$ has no clear physical
significance. The authors adopt the variant $t=N/2,$ where $N$ is the value at
which the $|\psi_{t}>$ basis is truncated, for reasons of rapid convergence of
numerical calculations. $\omega$ is also fixed numerically. The choice of a
shallow well, with $\alpha=2\lambda_{2}=-2,\ \beta=4\lambda_{4}=1,$ transforms
(17) in an equation with numerical coefficients, which gives a numerical value
for $r_{0}.$

Finally, the ground state eigenenergy in the presence of an electric field,
$\epsilon_{0}\left(  p\right)  ,$ is, for small and moderate values of the field:%

\begin{equation}
\epsilon_{0}\left(  p\right)  \simeq\epsilon_{0}\left(  0\right)  -\left\vert
a\right\vert p\tanh\omega p \label{18}%
\end{equation}
where the coefficients $a,\ \omega\ $are determined numerically. Their "exact"
value depends on the truncation of basis functions $|\psi_{t}>.$ In fact, this
formula of $\epsilon_{0}\left(  p\right)  $ is not rigorously deduced, but is
merely an interpolation between the "small" and "moderate" field cases. The
corectness of this ansatz could be evaluated according to its success of
giving reasonable predictions for the statistical physics of the
Ginzburg-Landau systems, through the transfer matrix method.

For large fields,%

\begin{equation}
\epsilon_{0}\left(  p\right)  \simeq A_{0}+B_{0}p^{4/3} \label{19}%
\end{equation}
with $A_{0},\ B_{0}\ -$ numerical constants. A similar - but more general -
result has been obtained by Bronzan and Sugar \cite{BrSg}, for the energy
levels of a potential $V(x)=A_{1}x+x^{4}.$ It is clear that, for large fields
(with Bronzan and Sugar's notations, for $A_{1}>15$), the effect of the
quadratic (harmonic) term of the potential is irrelevant, at least for the
leading terms.

Let us define what we mean here by a "large" field. In the absence of an
external field, the particle oscillates in a symmetric, double-well potential:%

\begin{equation}
V_{s}(x)=\frac{\alpha}{2}x^{2}+\frac{\beta}{4}x^{4} \label{20}%
\end{equation}
In the presence of an electric field, one gets a total, asymmetric, potential:%

\begin{equation}
V_{as}(x)=\frac{\alpha}{2}x^{2}+\frac{\beta}{4}x^{4}-px \label{21}%
\end{equation}

The "electric" term favors one of the two wells; this one becomes deeper,
while the other one - more shallow, disappearing when the electric field is
larger than a certain critical value $p_{c}.$ It is easy to find $p_{c}$ by
analyzing the extremum points of \ $V_{as}(x)$, which are the roots of the equation:%

\begin{equation}
\frac{dV_{as}(x)}{dx}=\beta x^{3}+\alpha x-p=0 \label{22}%
\end{equation}
where $\alpha<0,\ \beta>0,\ p>0.$ This equation may have one or three real
roots, according to the sign of the discrininant, proportional to the expression:%

\[
-\frac{\left\vert \alpha\right\vert ^{3}}{27\beta}+\frac{p^{4}}{4}
\]
which changes its sign at the critical value:%

\begin{equation}
p_{c}=\left(  \frac{4}{27}\frac{\left\vert \alpha\right\vert ^{3}}{\beta
}\right)  ^{1/4} \label{23}%
\end{equation}
If we presume a Landau-type temperature dependence of the parameter $\alpha, $
i.e. $\alpha=\alpha^{\prime}\left(  T-T_{c}^{MF}\right)  ,$ then the critical
field has also a $T-$dependence, of the form:%

\begin{equation}
p_{c}=p_{c0}\left\vert T-T_{c}^{MF}\right\vert ^{3/4} \label{24}%
\end{equation}
where we have used the standard notations of the Landau theory of phase
transitions (see for instance \cite{Chandra}).

\section{\bigskip The electrical polarization of the anharmonic chain}

The polarization is proportional to the average displacement of atoms along
the chain; according to (12), one obtains:%

\begin{equation}
\left\langle Q\right\rangle =\tanh\omega p\ +\ \frac{\omega p}{\cosh^{2}\omega
p},\ \ \ p<p_{c} \label{25}%
\end{equation}

For small fields,%

\begin{equation}
\left\langle Q\right\rangle =\omega p\left(  1+...\right)  \label{26}%
\end{equation}
and, comparing with the magnetization of the Ising model for small magnetic
fields \cite{Ising},%

\[
M=\beta He^{2\beta E_{1}}\ +\mathcal{\ O}\left(  \beta^{2}H^{2}\right)
\]
we can identify the coefficient $\omega$ as%

\begin{equation}
\omega\simeq\beta e^{2\beta cu_{0}^{2}} \label{27}%
\end{equation}
Qualitatively, (25) has a correct behaviour, in the sense that it is a
monotonically increasing function of $p$, which saturates asymptotically. The
same behaviour occurs in the 1D Ising model in an external field. Also, the
comparison with the Ising model recovers the temperature dependence of the
numerical constant $\omega,$ entering in the solutions obtained in \cite{Van}.
This recovery is essential in order to obtain the statistical mechanics of the
systems under scrutiny.

With (25), (27), we find for the susceptibility of the chain the following formula:%

\begin{equation}
\chi=\frac{2\omega}{\cosh^{2}\omega p}\left(  1-\frac{\omega p\tanh\omega
p}{\cosh\omega p}\right)  ,\ \ p<p_{c} \label{28}%
\end{equation}

\section{The spatial array of Ginzburg-Landau chains}

Let us consider a planar array of $N$ chains. The $n-$th chain $\left(  1\leq
n\leq N\right)  $\ is described by a Hamiltonian $H_{cl}^{(n)}$, where
$H_{cl}^{(n)}$ is obtained from $H_{cl},$ eq.(3), replacing $u\left(
x\right)  \rightarrow u_{n}\left(  x\right)  ,\ p\left(  x\right)  \rightarrow
p_{n}\left(  x\right)  .$ Cyclic boundary conditions are imposed, so that the
chains $n=1$ and $n=N+1$ coincide. The interaction between chains
$n,n^{\prime}$ is given by:%

\begin{equation}
H_{int}=\frac{1}{2}\sum_{n\neq n^{\prime}}D_{nn^{\prime}}\left[  u_{n}\left(
x\right)  -u_{n^{\prime}}\left(  x\right)  \right]  ^{2} \label{29}%
\end{equation}

In a mean field approach, the Hamiltonian (29) is replaced by:%

\begin{equation}
-2q_{i}D\left\langle u\right\rangle \sum_{n}u_{n}\left(  x\right)  ,\ \ i=2,3
\label{30}%
\end{equation}
where $q_{i}$ is the number of near neighbors, in $i$ dimensions; we shall
focus here, however, only on the 2D case. The evaluation of the statistical
physics of the 2D system reduces, throug the matrix transfer method, to the
study of a Schrodinger equation describing a quartic oscillator in an external
field $\Gamma\left\langle u\right\rangle $\ [3]:%

\begin{equation}
\left[  -\frac{1}{2m^{\ast}}\frac{d^{2}}{du^{2}}-\frac{1}{2}\left\vert
A\right\vert u^{2}+\frac{1}{4}Bu^{4}-\Gamma\left\langle u\right\rangle
u\right]  \psi_{n}\left(  u\right)  =\epsilon_{n}\psi_{n}\left(  u\right)
\label{31}%
\end{equation}

Measuring the energy in units $\left\vert V_{0}\right\vert =A^{2}/4B$ and
making the changes:%

\begin{equation}
u\rightarrow\left(  \frac{2\left\vert A\right\vert }{B}\right)  ^{1/2}%
u;\ \ \ \left(  m^{\ast}\right)  ^{-1/2}\rightarrow\left(  \frac{\left\vert
A\right\vert }{B}\right)  \left(  \frac{1}{2}\left\vert A\right\vert \right)
\mu;\ \ \ \Gamma\rightarrow\frac{1}{8}\left\vert A\right\vert \gamma\label{32}%
\end{equation}
we get the following eigenvalue problem:%

\begin{equation}
\left[  -\frac{1}{2}\mu\frac{d^{2}}{du^{2}}-4u^{2}+4u^{4}-\gamma\left\langle
u\right\rangle u\right]  \psi_{n}\left(  u\right)  =\epsilon_{n}\psi
_{n}\left(  u\right)  \label{33}%
\end{equation}
where:%

\begin{equation}
u_{0}=2^{-1/2},\ \ \ \mu=\frac{1}{4}\left(  \frac{k_{B}T}{V_{0}}\right)
\left(  \frac{\left\vert A\right\vert }{C}\right)  ^{1/2};\ \ \ \mu\sim T
\label{34}%
\end{equation}
In the classical paper of Bishop and Krumhansl [BK], the eigenvalue problem is
solved numerically. Our contribution, to this point of the problem, is the
following: using the analytical solution (anzatz) of Van der Straeten and
Naudts \cite{Van}, we shall obtain analytical formulae for the critical
temperature of the 2D transition, and the critical behaviour of the planar
array of chains.

Let us now outline our approach to the 2- and 3D problems and describe the
main results obtained.

\textbf{The mean field equation}

Replacing in (11)%

\begin{equation}
p=\gamma\left\langle u\right\rangle \label{35}%
\end{equation}
we get the self-consistent \ equation:%

\begin{equation}
bx=x+\tanh x-x\tanh^{2}x,\ \ b=\frac{1}{\omega\gamma},\ \ x=\omega
\gamma\left\langle u\right\rangle \label{36}%
\end{equation}
For $b<2$, the line described by the left hand side of the equation has a
non-zero intersection with the curve described by tht right hand side,
consequently a value $\left\langle u\right\rangle \neq0$ does exist; so, an
ordered state appears, meaning that the 2D system displays a phase transition.
Consequently, $b=2$ reprezints the critical condition, giving the expression
of the transition temperature.%

\[
\]

\bigskip\textbf{Evaluation of the critical temperature }$T_{c}$

According to eq. (27),%

\begin{equation}
\omega=\frac{1}{k_{B}T}\exp\left(  \frac{2}{k_{B}T}cu_{0}^{2}\right)
\label{37}%
\end{equation}
and the critical condition can be written as:%

\begin{equation}
\omega=\frac{1}{k_{B}T}\exp\left(  \frac{2}{k_{B}T}cu_{0}^{2}\right)
\label{38}%
\end{equation}
Putting%

\begin{equation}
\frac{2}{k_{B}T_{c}}cu_{0}^{2}=\xi\label{39}%
\end{equation}
it takes the form of the transcendental Lambert-Euler equation:%

\begin{equation}
\xi e^{\xi}=\frac{cu_{0}^{2}}{\gamma} \label{40}%
\end{equation}
Its solution is given by the Lambert function $W$ \cite{W} :%

\begin{equation}
\xi=W\left(  \frac{cu_{0}^{2}}{\gamma}\right)  \label{41}%
\end{equation}
The Lambert function has the following behaviour, for small, respectively
large value of its argument:%

\begin{equation}
W(\xi)\sim%
%TCIMACRO{\QATOPD{\{}{.}{\xi,\ \xi\sim0}{\ln\xi,\ \xi\rightarrow\infty} }%
%BeginExpansion
\genfrac{\{}{.}{0pt}{}{\xi,\ \xi\sim0}{\ln\xi,\ \xi\rightarrow\infty}
%EndExpansion
\label{42}%
\end{equation}
So, for small values of the inter-chain coupling, proportional to $\gamma,$%

\begin{equation}
T_{c}\sim\frac{1}{\left\vert \ln\gamma\right\vert } \label{43}%
\end{equation}
This result coincides to that obtained by Scalapino, Imry and Pincus
\cite{scal.im.pinc}, using a different approach. For moderate values of the coupling,%

\begin{equation}
T_{c}\sim\gamma\label{44}%
\end{equation}
The formulae (43), (44) confirm, qualitatively, the behaviour of $T_{c},$
obtaind numerically in \cite{BK}. Actually, it corrects their interpretation
for small $\gamma$ (considered by these authors to be exponential), and
provides an analytical expression for $T_{c}$, valid for any value of the
inter-chain coupling.%

\[
\]

\textbf{The critical behaviour of the order parameter}

In order to obtain the critical behaviour of the order parameter, we have to
examine the self-consistent condition (36) for temperatures close to $T_{c},$
where, putting $\omega(T_{c})=\omega_{c},$%

\begin{equation}
\omega=\omega_{c}+\Delta\omega,\ \ \ \Delta\omega\ll\omega_{c} \label{45}%
\end{equation}
So,%

\begin{equation}
\frac{1}{\omega\gamma}=2-2\gamma\Delta\omega\label{46}%
\end{equation}
Expanding the hyperbolic functions near the origin, we get:%

\begin{equation}
\left\langle u\right\rangle \sim\left\vert T-T_{c}\right\vert ^{1/2}
\label{47}%
\end{equation}
showing a mean field critical behaviour, as we can expect, taking into account
the methodology itself. However, our approach is interesting, due to the fact
that it also allows a simple, analytical determination of the coefficient of
$\left\vert T-T_{c}\right\vert ^{1/2}.$

\bigskip\textbf{The expression of the electrical susceptibility}

In the presence of an external electric field, the Hamitonian obtained through
the transfer matrix approach is:%

\begin{equation}
H_{q}^{(\text{modif})}=-\frac{1}{2m^{\ast}}\frac{d^{2}}{du^{2}}+\frac{1}%
{2}Au^{2}+\frac{1}{4}Bu^{4}-\Pi u,\ \ \ \Pi=\mathcal{E+\gamma}\left\langle
u\right\rangle \label{48}%
\end{equation}
So, the ground state energy becomes:%

\begin{equation}
\epsilon_{0}(\Pi)=\epsilon_{0}(0)-\left\vert a\right\vert \Pi\tanh\omega
\Pi\label{49}%
\end{equation}
and the self-consistency condition takes the form:%

\begin{equation}
\left\langle u\right\rangle =\tanh\left(  \mathcal{E+\gamma}\left\langle
u\right\rangle \right)  +\omega\left(  \mathcal{E+\gamma}\left\langle
u\right\rangle \right)  \left[  1-\tanh^{2}\omega\left(  \mathcal{E+\gamma
}\left\langle u\right\rangle \right)  \right]  \label{50}%
\end{equation}
which can be also written as:%

\begin{equation}
\left(  b-1\right)  y-e_{0}=\tanh y-y\tanh^{2}y,\ \ y=\omega\left(
\mathcal{E+\gamma}\left\langle u\right\rangle \right)  ,\ \ e_{0}%
=-\frac{3\mathcal{E}}{4\Gamma} \label{51}%
\end{equation}
On this equation, one can easily see that the line defined by the left hand
side has always an intersection with the curve defined by the right side, in a
point having the abscise $x>0.$ Consequently, in the presence of an external
field, $\left\langle u\right\rangle \neq0,$ as expected. For small values of
$y,$ the self-consistency condition gives the following equation:%

\begin{equation}
\frac{4}{3}y^{3}+\left(  b-2\right)  y-e_{0}=0 \label{52}%
\end{equation}
The physically interesting regime is still that corresponding to $b<2,$ so, putting%

\[
\frac{\Delta T}{T_{c}}=\tau
\]
we get the solution:%

\begin{equation}
\left\langle u\right\rangle =\frac{1}{\omega\Gamma}\left(  \frac{3e_{0}}%
{8}\right)  ^{1/3}\left\{  \left[  1+\left(  1-\frac{8}{9e_{0}^{2}}\tau
^{3}\right)  ^{1/2}\right]  ^{1/3}+\left[  11\left(  1-\frac{8}{9e_{0}^{2}%
}\tau^{3}\right)  ^{1/2}\right]  ^{1/3}\right\}  -e_{0} \label{53}%
\end{equation}
The exponent of the critical isotherm can be obtained immediatly, putting
$\tau=0$ in the previous equation:%

\begin{equation}
\left\langle u\right\rangle =\frac{1}{\omega\Gamma}\left(  \frac{3\mathcal{E}%
}{8\Gamma}\right)  ^{1/3}+...,\ \ \left\langle u\right\rangle ^{3}%
\sim\mathcal{E\ } \label{54}%
\end{equation}
So, the exponent of the critical isotherm is $\delta=3,$ as expected from the
general theory [Stanley].

However, the critical behaviour of the susceptibility can be obtained simpler,
through the differentiation of (52). Keeping only the smallest terms, we have:%

\begin{equation}
\frac{dy}{de_{0}}=\frac{1}{b-2}\sim\frac{1}{\Delta T} \label{55}%
\end{equation}
As this report is proportional to the electrical susceptibility, we find for
its critical behaviour a Curie-type law:%

\begin{equation}
\chi\left(  T\simeq T_{c}\right)  \sim\frac{1}{T-T_{c}} \label{56}%
\end{equation}
The coefficient of the $\left(  T-T_{c}\right)  ^{-1}$ factor can be easily
obtained, using standard methods (see for instance \cite{Stanley}, Ch.6).

\bigskip

\section{Conclusions}

In this paper, we have applied the transfer matrix method in order to study
the statistical mechanics of 1D and 2D Ginzburg-Landau systems - chains of
anharmonic oscillators, or planar arrays of such chains, which can simultate
ultra-thin films of ferroelectric systems. The starting point of our approach
is a partially numeric / partially analytic expression for the ground state
energy of a quartic oscillator in an electric field, $\epsilon_{0}\left(
p\right)  $, recently obtained by Van der Straeten and Naudts. In fact, the
formula for $\epsilon_{0}\left(  p\right)  $ is mainly an ansatz than a
rigorous result. Comparing the predictions of this formula, via the transfer
matrix method, for the polarization of the Ginzburg-Landau chain in an
electric field, with the small-field limit of the exact formula of the
magnetization of the 1D Ising model in a magnetic field, the temperature
dependence of the ground state energy of the quartic oscillator in an external
field is restored. Due to this fact, the expression of the ground state energy
$\epsilon_{0}\left(  p\right)  $ can be used, through a mean field
approximation, in order to find the statistical mechanics of the 2D
Ginzburg-Landau system. The method gives an analytical formula of the critical
temperature of the 2D order/disorder transition, expressed as a Lambert
function of the inverse inter-chain coupling.

Through this approach, we are able not only to find the statistical mechanics
of 1D and 2D Ginzburg-Landau systems, but also to confirm the adequacy of the
ansatz made by Van der Straeten and Naudts while proposing a formula for the
ground state energy of a quartic oscillator in external field.

\bigskip

\begin{acknowledgement}
The author thanks to CNCSIS and UEFISCSU for funding the research presented in
this paper, through the project IDEI 2008 no. 953.
\end{acknowledgement}

\bigskip

\bigskip

\end{document}